\documentclass[
 reprint,
superscriptaddress,
 amsmath,amssymb,
 aps,
prb,
floatfix,
]{revtex4-2}
\usepackage{graphicx}
\usepackage{dcolumn}
\usepackage{bm}

\usepackage{hyperref}
\usepackage[normalem]{ulem} 
\usepackage[caption=false]{subfig}

\begin{document}


\title{Motility destabilizes an absorbing-state flock}

\author{Narayan Dutt Sharma}
\email{narayansd@iisc.ac.in}
\author{Subroto Mukerjee}
\email{smukerjee@iisc.ac.in}
\affiliation{Centre for Condensed Matter Theory, Department of Physics, Indian Institute of Science, Bangalore, 560 012, India}

\author{Chandan Dasgupta}
\email{cdgupta@iisc.ac.in}
\author{Sriram Ramaswamy}
\email{sriram@iisc.ac.in}
\affiliation{Centre for Condensed Matter Theory, Department of Physics, Indian Institute of Science, Bangalore, 560 012, India}
\affiliation{International Centre for Theoretical Sciences, TIFR, Bangalore 560 089, India}

\date{\today}

\begin{abstract}
Activity, when it takes the form of motility, is generally 
seen to promote order in many-body systems. Here we present a one-dimensional lattice model that, in the 
non-motile limit, exhibits absorbing ferromagnetic states. When activity is introduced through biased motility, these 
absorbing state are destabilised and the system
instead undergoes a 
transition from an ordered flock to a disordered state as the alignment strength is decreased. 
A finite-size scaling analysis of physical quantities reveals a continuous transition with critical exponents satisfying the 
hyperscaling relation in one dimension, providing quantitative evidence for the activity-induced disorder. 
\\

\end{abstract}

\maketitle




Self-propulsion is frequently seen to promote ordering in interacting many-body systems. Examples include the existence of long-range order in the Vicsek and Toner-Tu flocking models in two dimensions \cite{Vicsek1995,toner1995long, Chate2020} 
and condensation without attraction or depletion in non-aligning but persistently moving particles \cite{Tailleur2008,Fily2012,Redner2013,Cates2015,cates2025active}.

\begin{figure}
\includegraphics{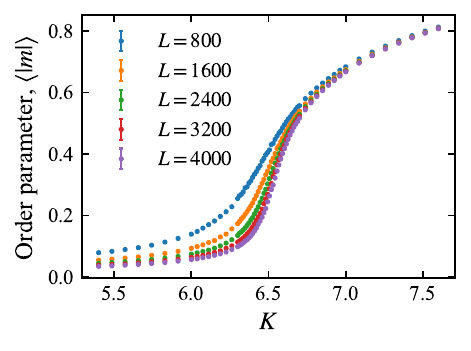}
\includegraphics{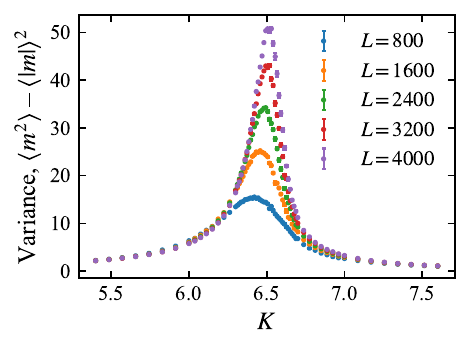}
\caption{\label{fig:Order Parameter and Variance} Order parameter and variance as functions of the alignment strength $K$ for a nearest neighbour interaction and $\rho=0.8$.}
\end{figure}

\begin{figure}
\includegraphics{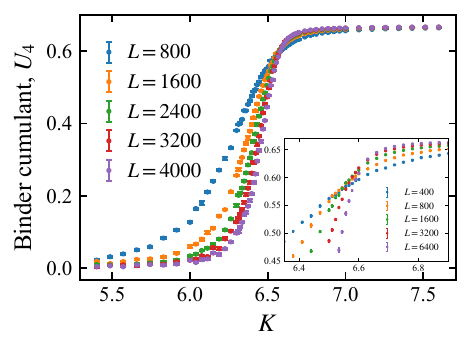}
\caption{\label{fig:Binder cumulant} Binder cumulant as a function of the alignment strength $K$ for a nearest-neighbour interaction and $\rho=0.8$. The inset shows a magnified view of the crossing region.  
}
\end{figure}

\begin{figure*}
    \includegraphics[scale=0.725]{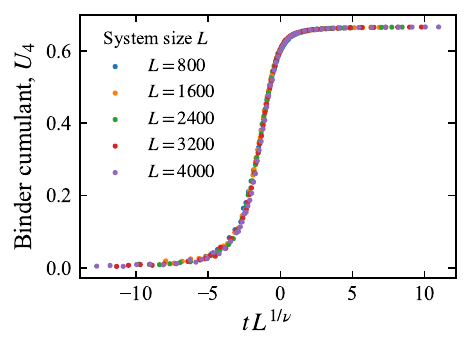}
    \includegraphics[scale=0.725]{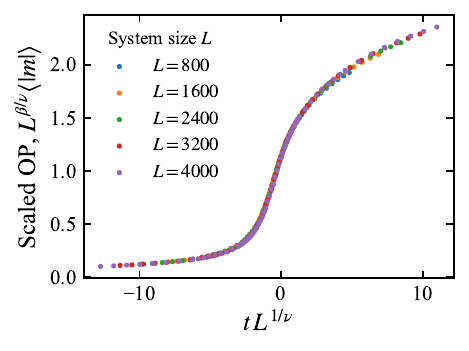}
    \includegraphics[scale=0.725]{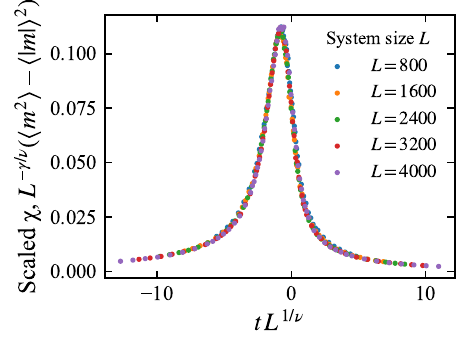}
    \caption{
        \label{fig:scaling_collapse}
        Finite-size scaling collapse for nearest-neighbour interactions at
        density $\rho=0.8$ using
        $K_c = 6.58$, $\nu = 1.95$, $\beta = 0.31$, and $\gamma = 1.29$.
        Left: Binder cumulant $U_4$.
        Middle: order parameter $\langle |m| \rangle$.
        Right: variance $\chi$ of the magnetization.
        $t$ is the reduced alignment, $t = (K-K_c)/K_c$.
    }
\end{figure*}

A distinct and broad class of nonequilibrium systems is characterized by absorbing states: configurations that, once reached, offer no way out.
These occur in directed percolation \cite{Hinrichsen2000, Odor2004}, reaction--diffusion processes \cite{Cardy1996, Elgart2006, Mussawisade1998} and spin models with zero-temperature dynamics \cite{Spirin2001, Bray1994}, in which parameters are tuned such that the dynamics eventually 
puts the system in an absorbing 
configuration.
We ask and answer the question: what does turning on activity do to an already existing absorbing state?
We introduce a one-dimensional Ising-like model whose spins are taken to point left ← or right → along the lattice, and whose single-spin-flip rule (1) has ferromagnetic absorbing states in its non-motile limit. When activity is introduced through motility in the direction of the spin, ferromagnetic order is destabilized for alignment interaction below a threshold value and the system undergoes a “deflocking” transition (Fig. \ref{fig:Order Parameter and Variance}), which we characterize via Binder cumulant crossings (Fig. \ref{fig:Binder cumulant}) and finite-size scaling (Fig. \ref{fig:scaling_collapse}).
The novel feature here is that the reference state in the absence of motility is ordered, and activity acts to destabilize the order.

Previous studies on one-dimensional active spin models report the formation of alternating flocks, where the 
direction of motion of the flock reverses stochastically on a time scale growing as $\ln L$ with system size 
\cite{OLoan1999,Raymond2006,Solon2013,Benvegnen2022}. A key distinction between those studies and the present work 
lies in the implementation of exclusion and the stochastic spin-flip process. As we show below, this 
modified microscopic dynamics leads not to alternating flocks but to a flocking phase transition 
in which activity destroys an otherwise absorbing ordered phase.

\begin{figure}
\includegraphics{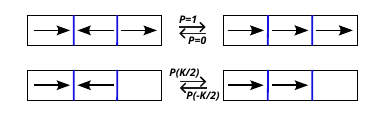}
\caption{\label{fig:Flip process} The asynchronous flip sweep for nearest neighbour interactions. The blue barrier indicates the flipped spin with its corresponding flip probability. 
The corresponding flip probabilities are obtained from Eq.~\ref{eq:flip probability}. In the figure, $P(x)$ denotes the flip probability evaluated at $x=K \sigma_i M_i$, giving the values $P(K/2)$ and $P(-K/2)$ shown on the arrows.}
\end{figure}

We first consider a one-dimensional model with one Ising spin $\sigma_i= \pm 1$ at each lattice sites labelled by $i = 1,2, ... L$ and periodic boundary conditions.
In the absence of any motility, the model evolves in time via a single spin-flip rule given by
\begin{eqnarray}
    P_{\sigma_i\rightarrow \bar{\sigma_i}} = \dfrac{e^{-K \sigma_i M_i}-e^{-K}}{e^{K} - e^{-K}}
    \label{eq:flip probability}
\end{eqnarray}
where $K$ is the alignment strength 
and $M_i=\tfrac{1}{2R} \sum_{j:0<|i-j|\leq R} \sigma_j$ is the normalized molecular field at site $i$ due to 
$R$ neighbouring sites 
on each side. 
Unlike the conventional Metropolis update \cite{Metropolis1953}, this spin-flip rule ensures a vanishing flip probability when the local molecular field is fully aligned with the onsite spin. The exponential dependence on the local alignment field is reminiscent of the rule used in Ref.~\cite{Khasseh2025}, but the present form is normalized so as to enforce absorbing aligned states. The motivation for adopting this form is to focus on the nonequilibrium ordering dynamics in the absence of bulk thermal fluctuations. In particular, the dynamics suppresses stochastic flips in locally ordered regions, thereby promoting the stability of globally aligned configurations.
Hence, by construction when the molecular field $M_i=\pm1$, that is, all neighbours are up or down, the flip probability is $0$ ($1$) if the onsite
spin is aligned (anti-aligned) as shown in the top image of Fig.~\ref{fig:Flip process}. 

In the nearest-neighbour limit ($R=1$), 
spin-flip probabilities take the deterministic values $P=0$ or $P=1$ whenever the local field is nonzero, as in zero-temperature Glauber dynamics for the kinetic Ising model \cite{Menyhard1994}. The two dynamics differ when the local field vanishes: in Glauber dynamics $P=1/2$, whereas in our model $P=(1-e^{-K})/(e^{K}-e^{-K})$, which approaches $1/2$ only in the limit $K \to 0$.
For any $K$ and any finite interaction range $R$, the dynamics relaxes the system to a fully ordered all-right or all-left configuration (corresponding to all-up or all-down states in the Ising representation), which once reached, is never changed. We therefore refer to these ferromagnetic configurations as absorbing states. 

We now relax the constraint of each site being occupied by a single spin and allow the total number of spins to be $N=L\rho$ with $0<\rho<2$. The spins adhere to the same species 
exclusion rule, so, at most one spin with $\sigma=+1$ (right) and one with $\sigma=-1$ (left) can occupy the same site. The
onsite magnetization $m_i$, defined as the sum of the spins at site $i$, can then take the values
$-1, 0, +1$ where $m_i=0$ can imply a vacancy or double occupancy, which can only be distinguished if the local density is specified. The same species exclusion rule implies that only spins at singly occupied sites can flip. The spin-flip rule is still given by Eq.~\ref{eq:flip probability} with $\sigma_j$ in the definition of the normalized molecular field $M_i$ replaced by the local magnetization $m_j$.  
Note that any state with no singly occupied sites is also a stationary state, but unlike the fully aligned states does not have a basin of attraction in 
configuration space, and hence is not an absorbing state.

\begin{figure}
\includegraphics{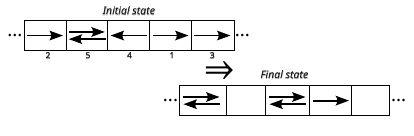}
\caption{\label{fig:Jump process} The asynchronous jump sweep. The number shows the order in which the moves are made. Both spins at a doubly occupied site jump in a single move if allowed to do so by the occupancy of its neighbours.}
\end{figure}

We now supplement this spin flip rule with biased motility. 
Interestingly, the motility prevents the system from relaxing into the ferromagnetic absorbing states and instead leads to a phase transition from an ordered flock to a disordered state 
as the alignment strength $K$ is varied.
The dynamics is implemented numerically as a two-step asynchronous process: an asynchronous hopping sweep, as shown in Fig.~\ref{fig:Jump process}, followed by an asynchronous spin-flip sweep. 
Fig.~\ref{fig:Flip process} illustrates the local spin-flip rule used during the latter. 
In each sweep, we generate a random permutation of lattice sites and update the sites in that order. 
We note that 
the transition is update-dependent, a point we return to later. In particular, the transition vanishes when a parallel synchronous update is used. Such update dependence is well known, e.g., for the steady-state properties of one-dimensional driven systems like the TASEP models
~\cite{Rajewsky1998, Wolki2006}.

We work at fixed global density $\rho$  and use the alignment strength $K$ as the control parameter.
The flocking order is quantified by the global magnetization
\begin{eqnarray}
    m = \frac{1}{N}\sum_{i=1}^L m_i,
\end{eqnarray}
whose steady state absolute value $\langle |m| \rangle$ distinguishes an ordered flocking phase ($m\neq 0$) from a disordered phase ($m = 0$) in the thermodynamic limit.
We further compute the variance, $\chi=N (\langle m^2\rangle -\langle|m|\rangle^2)$ and Binder cumulant, $U_4=1-{\langle m^4\rangle}/{3\langle m^2 \rangle^2}$ 
to locate and characterize the phase transition via finite size scaling \cite{Binder1988}. Fig.~\ref{fig:Binder cumulant} shows the 
Binder cumulant plot and Fig.~\ref{fig:Order Parameter and Variance} shows the order parameter and its variance for the system with
nearest-neighbour interaction and density $\rho=0.8$. Unless stated otherwise, all data shown henceforth is for this case.

We simulate systems of size up to $L=4000$ with periodic boundary conditions, discarding initial transients and computing steady-state averages over independent realizations.
As $K$ is increased, the steady-state magnetization changes smoothly from a value characteristic of the 
disordered regime with small $\langle |m|\rangle$ to a large value characteristic of an ordered flocking regime,
while $\chi$ develops a peak and $U_4$ exhibits a common crossing point, indicating a continuous order--disorder 
phase transition.

To locate the critical point $K_c$, we use the crossing of the Binder cumulant $U_4(K, L)$. As shown in Fig.~\ref{fig:Binder cumulant},
curves of different system sizes intersect at a common point $K\simeq6.58$ providing an estimate of the critical alignment strength.
Close to $K_c$, the Binder cumulant, the order parameter and its variance are expected to follow the finite-size scaling forms
\begin{eqnarray*}
    U_4 &=& f_U(t L^{1/\nu})\\
    \langle |m| \rangle &=& L^{-\beta/\nu}f_m(t L^{1/\nu})\\
    \chi &=& L^{\gamma/\nu}f_\chi(t L^{1/\nu})
    \label{eq:Binder scaling}
\end{eqnarray*}
where $t= \tfrac{(K-K_c)}{K_c}$  and $\nu, \beta \text{ and } \gamma$ are the critical exponents. To extract $K_c$ and the critical exponents together with their respective
uncertainties, we employ the scaling-collapse procedure of Bhattacharjee and Seno~\cite{Bhattacharjee2001},
which minimizes a relative, rather than absolute, penalty function. We first determine $K_c$ and $1/\nu$ from the Binder 
cumulant data collapse, and subsequently extract $\beta/\nu$ and $\gamma/\nu$ from the scaling collapse of
the magnetization and susceptibility, respectively. 
The details of this procedure are given in Appendix~\ref{app:Numerical_Procedure}.
The resulting critical exponents are
\[
\begin{aligned}
\nu &= 1.95 \pm 0.18, \\
\beta &= 0.31 \pm 0.05,\\
\gamma &= 1.29 \pm 0.13
\end{aligned}
\]

The corresponding scaling collapses of $U_4$, $\langle |m| \rangle$,
and $\chi$ for $\rho=0.8$ in the nearest neighbour case are shown in Fig.~\ref{fig:scaling_collapse}.

The extracted exponents are internally consistent and satisfy the hyperscaling relation
\[
    2\beta/\nu + \gamma/\nu = d,
\]
which yields $2\beta/\nu + \gamma/\nu = 0.98 \pm 0.07$, in excellent agreement with the spatial dimension $d=1$. 
This provides quantitative evidence that the flocking transition that we observe is a genuine continuous phase transition. 
Representative videos of the (quenched) asynchronous dynamics are available at Ref. \cite{SupplementaryVideos}.

\begin{figure}
\includegraphics{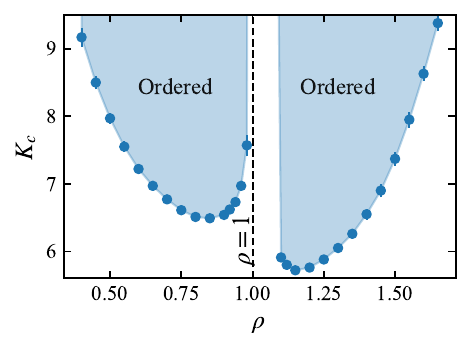}
\caption{\label{fig:phase_diagram_R1}
        Phase diagram $K_c(\rho)$ for nearest-neighbour interactions ($R=1$). As $\rho \to 1$, no finite critical alignment strength is observed, effectively $K_c \to \infty$. Near $\rho=1$, strong fluctuations and pronounced finite-size effects make the precise identification of the transition increasingly difficult.} 
\end{figure}

An important finding, for the nearest-neighbour interactions considered so far, is that 
the density $\rho=1$ is special: it yields 
no
flocking transition at any finite $K$, i.e., $\rho=1$ corresponds to the limit $K_c \to \infty$. The resulting phase diagram $K_c(\rho)$ 
for $R=1$ is shown in Fig.~\ref{fig:phase_diagram_R1}. For larger interaction ranges $R>1$, by contrast, the point $\rho=1$ is no longer
singular; instead $K_c(\rho)$ develops a minimum near $\rho=1$ as shown in Appendix \ref{app:Phase_diagram_R2}.

In the purely relaxational limit without motility, the dynamics for $R=1$ resembles zero-temperature Glauber dynamics: 
domain walls  perform a random walk and can only annihilate but never be created. The fully ordered ferromagnetic states are therefore 
strictly absorbing. Introducing biased motility qualitatively changes this picture. Due to the spins drifting in their respective directions, 
the system explores a much larger variety of local configurations (neighbours) than in the non-motile case, opening new ``channels" for disordering.

The same-species exclusion constraint forces doubly occupied sites to have $m_i=0$, which thus act as mobile sources of domain 
walls. Although spin flips are not attempted at such sites, their presence strongly reduces the local molecular field $M_i$ experienced by 
neighbouring spins.
In particular, for nearest neighbour interaction ($R=1$), the molecular field generated by a neighbouring
vacancy and doubly occupied site vanishes exactly, $M_i=0$. Consequently, the flip probability becomes 
independent of the spin orientation, effectively promoting unbiased local flips and generating new domains.
Motility thus opens additional channels for defect conversion, not present in the non-motile case, whose rate competes with the alignment-driven
annihilation of domain walls. Below $K<K_c$, the defect-creation process induced by motility dominates, sustaining a fluctuating disordered
flock, whereas for $K>K_c$ the alignment term prevails and the system orders.
For nearest-neighbour interactions and density $\rho=1$, motility-mediated collisions are particularly frequent, 
leading to persistent configurations where
$M_i=0$ occurs repeatedly, thereby preventing long-range order for any finite $K$. 
This mechanism explains the phase diagram for the nearest-neighbour case Fig.~\ref{fig:phase_diagram_R1} where no finite $K_c$ is observed at $\rho=1$, despite the presence of an absorbing state in the non-motile limit. 
For $R>1$, however, the molecular field is nonzero on average, and a transition is recovered at $\rho=1$, with $K_c$ exhibiting a minimum near this density. 
If $\rho$ is sufficiently less than unity so that the average spacing between singly occupied sites is
greater than $R$, 
 these sites with $m_i \neq 0$ 
 are so distant from each other that they cannot force spin alignment. Thus, in the absence of motility, the spins at these sites keep flipping back and forth and there is no stationary state and hence also no absorbing state. This implies that $K_c \rightarrow \infty$ for sufficiently small values of $\rho$ for a given $R$, which is consistent with the rapid upturn in the phase boundary at $\rho \approx 0.5$ seen in Fig.~\ref{fig:phase_diagram_R1}. A similar argument holds for $\rho$ sufficiently greater than unity with double occupancy being responsible for the large fraction of $m_i=0$ sites.

Considering that the maximum global magnetization achievable by the system is $m = \rho$ for $0< \rho \leq 1$ and $m = 2-\rho$ for $1\leq \rho < 2$, one might expect the model to be symmetric about $\rho=1$ under the transformation $\rho \rightarrow (2-\rho)$, $m \rightarrow -m$ and $x \rightarrow -x$. This symmetry also appears in the deterministic coarse grained equations derived from the microscopic dynamics 
(Appendix ~\ref{app:Coarse grained (hydrodynamic) equations}). At the microscopic level, however, configurations 
with zero local magnetization are not equivalent: a site with $m=0$ may correspond either to a vacancy or to a 
doubly occupied site. While a vacancy suppresses fluctuations due to the absence of particles, a doubly occupied 
site can generate two outgoing spins and thus acts as a local source of fluctuations. This distinction is not 
captured by the deterministic coarse-grained description when only the magnetization field is retained.
Moreover, due to the asynchronous (random sequential) update, the microscopic dynamics is not invariant under 
the transformation mentioned above: the evolution of a given configuration and that of its symmetry-transformed counterpart are not equivalent. This further reinforces the absence of an exact $\rho \rightarrow 2-\rho$ symmetry at the 
microscopic level.
This asymmetry is also evident in
Figs.~\ref{fig:phase_diagram_R1} and ~\ref{fig:phase_diagram_R2}, which are not symmetric about $\rho=1$. So, despite the apparent symmetry at the
coarse-grained deterministic level, the microscopic dynamics does not possess an exact $\rho \rightarrow 2-\rho$ symmetry.
Since microscopic
fluctuations are suppressed in low density regions~\cite{Dean1996}, a density dependent noise term, which would remove the symmetry, is expected on physical grounds. 
Incorporating such a 
term may resolve the observed asymmetry which we defer to future work.

Interestingly, under the synchronous (parallel) update, for $\rho=1$ (for $R=1$, and more generally for any 
finite $R$) the system can still reach absorbing states. In this case, double occupancy is effectively suppressed by the 
update rule, and the system evolves towards configurations similar to the non-motile limit, where dynamics occurs 
only at domain walls separating right and left oriented regions. This restriction confines the dynamics to domain wall motion, leading the system to absorbing configurations.
At very low values of $K$, the 
system admits four absorbing states: two ferromagnetic and two antiferromagnetic. At higher $K$, the dynamics 
preferentially selects the ferromagnetic absorbing states, although the time required to reach these states 
increases with increasing $K$, reflecting the associated change in flip probability. Representative videos for the synchronous dynamics are available at Ref. \cite{SupplementaryVideos}.

This behavior highlights the crucial role of motility in generating the fluctuating phase observed under 
asynchronous updates. The update scheme for motility qualitatively alters the accessible configurations. For 
example, starting from a configuration with a small number of domain walls, synchronous dynamics allows only 
their diffusive motion, whereas asynchronous dynamics enables spins to penetrate ordered regions, generating 
additional defects and sustaining fluctuations. The flip dynamics, however, is insensitive to the update scheme. 
A more detailed analysis of the parallel dynamics is provided in Appendix~\ref{app:parallel_update}.

In this work, we have investigated a one-dimensional lattice model with biased dynamics and local interactions, subject to a constraint
that enforces the conservation of particle density. In the absence of motility, the model reaches an absorbing state, whereas with motility a non-trivial steady state is achieved. Using extensive numerical simulations and a finite size scaling analysis, we have characterized the transition to this steady-state as an order-disorder flocking transition controlled by an alignment parameter and have extracted the  critical exponents 
associated with the transition. We find that the critical exponents satisfy the hyperscaling relation for $d=1$.
Although the exponent estimates exhibit mild density-dependent variations, we observe no trend beyond statistical uncertainty, supporting the conclusion that the transition belongs to a single universality class (see Appendix \ref{app:Phase_diagram_R2}).

We note that a recent study has proposed a corresponding classical transition in a one-dimensional quantum flocking model ~\cite{Khasseh2025}.
However, the microscopic dynamics of that model differ fundamentally from ours. In particular, the spin-flip rule employed here 
allows for the formation of absorbing ferromagnetic states, a feature absent in the quantum model. Consequently, the transition reported
here arises from the activity-induced destabilisation of an absorbing phase, highlighting a different mechanism in one dimension.

The emergence of order in this system can be understood as a consequence of the competition between biased transport and local spin-alignment processes.
Double occupancy plays a non-trivial role here by suppressing local magnetization, thereby modifying the effective interaction experienced by the spin. This mechanism may explain the absence
of critical behavior in the nearest-neighbour model with $\rho=1$. The particle constraint represents a key distinguishing  feature of the present model  compared to previously studied classical one-dimensional biased lattice models, suggesting that it is 
a crucial ingredient underlying the observed dynamics. Overall, our results demonstrate that activity can destabilize an absorbing ordered phase, leading to a phase transition in one dimension.

\textit{Acknowledgements:}
SR thanks the ANRF, India, for support in the form of a J C Bose National Fellowship (till Nov 2025) and a National Science Chair thereafter. CD acknowledges support from the National Academy of Sciences, India in the form of a Senior Scientist Fellowship.

\bibliography{aps}

\clearpage
\onecolumngrid

\newpage

\appendix

\begin{center}
    \textbf{Supplementary material}
\end{center}

\section{Coarse grained equations}
\label{app:Coarse grained (hydrodynamic) equations}
The microscopic equation for the dynamics is as follows. Suppose $\epsilon$ indicates the strength of the bias of the spin and $s=\pm$, then the microscopic equation 
can be written as-

\begin{eqnarray*}
\partial_t n^{s}_i = 
\dfrac{(1+s\epsilon)}{2} n_{i-1}^s(1-n_{i}^s) - \dfrac{(1+s\epsilon)}{2}n_i^s (1-n_{i+1}^s) 
- \dfrac{(1-s\epsilon)}{2} n_i^s (1-n_{i-1}^s) + \\
\dfrac{(1-s\epsilon)}{2}n_{i+1}^s (1-n_i^s) 
+ \Gamma \!\left[n_i^{\bar{s}}(1-n_i^s) P(\bar{s},K)- n_i^s(1-n_i^{\bar{s}}) P(s, K) \right]
\end{eqnarray*}
where $\epsilon=0$ translates to pure diffusion and $\epsilon=1$ translates to pure bias. Here, $n_i^{\pm} \in \{0, 1\}$. Now, define 
$\rho_i = \langle n_i^+ + n_i^- \rangle$ and $m_i = \langle n_i^+ - n_i^- \rangle$. 
Define mean fields, $p_i^{\pm}=\langle n_i^\pm\rangle = \tfrac{1}{2}(\rho_i \pm m_i)$.  
The transformation leads to the coarse grained PDEs -

\begin{eqnarray}
    \partial_t\rho = \tfrac{1}{2} \partial^2_x \rho + \epsilon \partial_x [(\rho-1)m]
\end{eqnarray}

\begin{align}
    \partial_t m = \tfrac{1}{2}  \partial^2_x m -\epsilon(1-\rho)\partial_x \rho + \epsilon m \partial_x m + \mathcal{F}_{\mathrm{flip}} 
\end{align}

where $\mathcal{F}_{\mathrm{flip}}$ (in terms of neighbourhood field, $M_i = \dfrac{1}{2R}\sum_{j:0<|i-j|<R} m_j$) is

\begin{eqnarray}
    \tfrac{\Gamma}{\sinh{K}} \Big{[}  \tfrac{1}{2}(2\rho - \rho^2) \sinh{KM} - m\cosh{KM}+ me^{-K}  + \tfrac{m^2}{2} \sinh{KM} \Big{]}
    \label{eq:F_flip}
\end{eqnarray}

The homogeneous solution is given $\mathcal{F}_{flip}=0$. Assuming mean field, the transcendental equation is

\begin{eqnarray}
    m = \dfrac{\tfrac{\rho}{2} (2-\rho)\sinh{Km}}{\cosh{Km}-e^{-K}-\tfrac{m}{2}\sinh{Km}}
\end{eqnarray}

Solving the above equation reveals that for $0<\rho<2$ and $\rho \neq 1$, the only solution of the equation is $m=0$ at low $K$ and it 
has three solution at high value of $K$ which are $m=0, \pm m_0$. Of these, only $m=\pm m_0$ is stable. 
For $\rho=1$, the equation always has three solutions, however, at low $K$ only $m=0$ is stable and the other two become stable at 
high sufficiently large $K$. It is worth noting that the $R$ enters only through the normalization of molecular field and the equations become independent of it when under the mean field assumption.

The equation can also be cast into Ginzburg-Landau power form. 
Assuming a smooth magnetization field near the phase transition, the neighbourhood field $M=\sum_r w_r m_r$, $\sum w_r=1$  can be expanded as  
\begin{eqnarray}
    M \approx m + \nu \partial_x^2 m + \mathcal{O}(m^3)
    \label{eq:M approx}
\end{eqnarray}

Let $\mathcal{S}(\rho)=\tfrac{\rho}{2}(2-\rho)$. Substituting ~\ref{eq:M approx} back in ~\ref{eq:F_flip} gives 

\begin{eqnarray*}
    \dfrac{\Gamma}{\sinh{K}}[(e^{-K} - 1 + \mathcal{S} K)m + 1/2(K^3\mathcal{S}/3-K^2+K)m^3]
    + \dfrac{\mathcal{S}\Gamma}{\sinh{K}} \nu \partial^2_x m 
\end{eqnarray*}

In the numerics, since each site is flipped once every time step $\Gamma=1$. The full m equation then becomes 
\begin{eqnarray}
    \partial_t m = \tilde{D}_m \partial_x^2m - \epsilon(1-\rho) \partial_x \rho + \epsilon m \partial_x m + 
    a(K,\rho)m + b(K,\rho)m^3
\end{eqnarray}

where effective diffusion coefficient is  $\tilde{D}_m = 1/2 + \dfrac{\mathcal{S}\Gamma}{\sinh{K}} \nu$, 
and the landau coefficients are 
\begin{eqnarray*}
a(K, \rho)=\dfrac{1}{\sinh{K}} \left(e^{-K}-1+\mathcal{S}K\right), \quad 
b(K, \rho)=\dfrac{1}{2\sinh{K}}\left(\tfrac{K^3\mathcal{S}}{3} -K^2+K\right).
\end{eqnarray*}

For $\rho=0$ and $\rho=2$, all coefficients remain negative for all $K$, consistent with the absence of ordering. 
For intermediate densities, both coefficients vary with $K$ and can change sign at different values of $K$: they 
are negative at small $K$ and become positive at sufficiently large $K$. This behavior indicates that the simple 
Landau expansion does not provide a controlled description of the transition across the full parameter range, and 
therefore does not capture the full behavior of the system.

\section{Numerical Procedure to extract the exponents}
\label{app:Numerical_Procedure}

The finite-size scaling forms for the Binder cumulant, order parameter, and susceptibility are given by
\begin{eqnarray}
    U_4 &=& f_U(tL^{1/\nu}), \nonumber \\
    \langle m \rangle &=& L^{-\beta/\nu}f_m(t L^{1/\nu}), \nonumber \\
    \chi &=& L^{\gamma/\nu}f_\chi(t L^{1/\nu})
    \label{eq:Binder_scaling}
\end{eqnarray}

where $t = \tfrac{K-K_c}{K_c}$ is the reduced control parameter.
Following Ref.~\cite{Bhattacharjee2001}, we require that all rescaled data collapse onto a single universal curve.
To quantify the quality of collapse, we construct a penalty function by taking one dataset as a reference curve, interpolating it, and measuring the 
deviation of all other datasets from this reference.
The modified penalty function is defined as-

\begin{eqnarray}
\label{eq:Penalty function}
    P_b = \dfrac{1}{\mathcal{N}_{\mathrm{over}}} \sum_p \sum_{j\neq p}\sum_{i \in \mathrm{over}} 
    \dfrac{|L_j^{-\Delta} \mathcal{O}_{i,j} - \mathcal{E}_p(L_j^{1/\nu} t_{ij})|}
    {L_j^{-\Delta} \mathcal{O}_{i,j} + \mathcal{E}_p(L_j^{1/\nu} t_{ij})},
\end{eqnarray}

where $\Delta$ is the scaling exponent associated with the observable $\mathcal{O}$($\Delta=0$ for Binder cumulant), and $\mathcal{E}_p(x)$ is the interpolating function constructed from the $p^{\mathrm{th}}$ dataset.
The first sum runs over all choices of reference curves, the second over all other datasets, and the third over all overlapping points in the rescaled 
variable. The normalization factor $\mathcal{N}_{\mathrm{over}}$ denotes the total number of such comparisons.

The use of the relative deviation in Eq.~\ref{eq:Penalty function} ensures that the penalty function remains well-behaved during optimization. In particular, using an absolute deviation alone can lead to a monotonic increase in the penalty as $\Delta$ is varied, preventing the existence of a well-defined minimum.
We minimize $P_b$ using the \textit{scipy.optimize} package with the Nelder--Mead method, which is a derivative-free simplex-based optimization algorithm.

The procedure to extract $K_c$, $\nu$, $\beta$, and $\gamma$ is as follows. The critical coupling $K_c$ and the correlation-length exponent $\nu$ are 
obtained from the finite-size scaling collapse of the Binder cumulant, while the ratios $\beta/\nu$ and $\gamma/\nu$ are extracted from the corresponding
collapses of the magnetization and susceptibility, respectively.
The Binder cumulant is optimized independently for each density to obtain estimates of $\nu$ and $K_c$. The data points are restricted to the critical region defined by $|t L^{1/\nu}| < X_{\mathrm{max}}$. For $R=1$, we choose $X_{\mathrm{max}}=1.5$, which ensures that a sufficient number of points are included in the fit while remaining within the scaling regime. The average and standard deviation over densities provide the estimates and associated uncertainties for universal exponent $\nu$.
Using the extracted values of $\nu$ and the corresponding $K_c$, the magnetization and susceptibility collapses are used to compute the ratios $\beta/\nu$ and $\gamma/\nu$ at each density. The averages and standard deviations of these quantities, over density, are then used to obtain $\beta$ and $\gamma$ and their uncertainties. Due to the dependence on $\nu$, the error bars on $\beta$ and $\gamma$ are obtained via error propagation:

\begin{eqnarray*}
    \Delta \beta &=& \sqrt{((\tfrac{\beta}{\nu})\Delta \nu)^2 +(\nu \Delta (\tfrac{\beta}{\nu}))^2}\\
    \Delta \gamma &=& \sqrt{((\tfrac{\gamma}{\nu})\Delta \nu)^2 +(\nu \Delta (\tfrac{\gamma}{\nu}))^2}
\end{eqnarray*}

\begin{figure}
    \label{fig:exponent_variation}
    \includegraphics[scale=0.725]{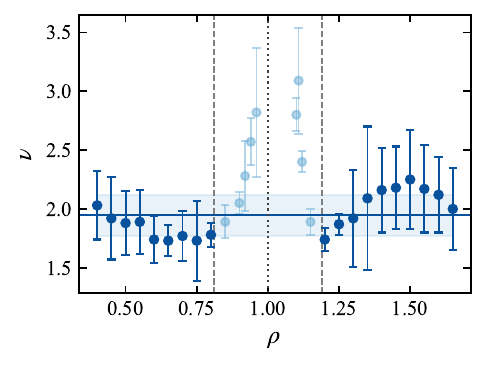}
    \includegraphics[scale=0.725]{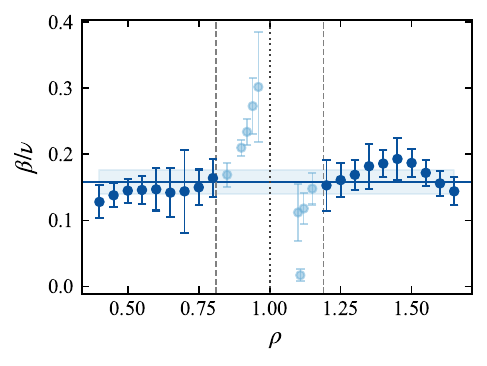}
    \includegraphics[scale=0.725]{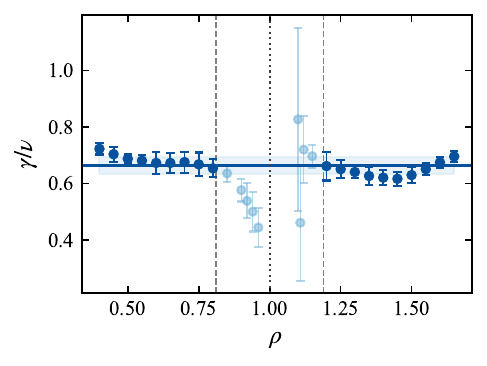}
    \caption{
        \label{fig:exponent_variation}
        The exponent variation as a function of density for R=1. The points in the region $\rho \in (0.81, 1.19)$ are excluded because of the singularity at $\rho=1$. Top left: Variation of $\nu$, Top right: Variation of $\beta/\nu$, Bottom: Variation of $\gamma/\nu$.
    }
\end{figure}

An independent estimate of the uncertainties can be obtained from the curvature of the penalty function near its minimum, following Ref.~\cite{Bhattacharjee2001}. Assuming the penalty function is locally quadratic, we expand $\ln P_b$ around the minimum along each parameter direction, yielding

\begin{eqnarray}
    \label{eq:Error_estimate}
    \Delta K_c &=& \eta K_c \left[ 2 \ln{\dfrac{P_b(K_c \pm \eta K_c \,|\, \nu)}{P_b(K_c)}} \right]^{-1/2}, \nonumber\\
    \Delta \nu &=& \eta \nu \left[ 2 \ln{\dfrac{P_b(\nu \pm \eta \nu \,|\, K_c)}{P_b(\nu)}} \right]^{-1/2}, \nonumber\\
    \Delta \beta &=& \eta \beta \left[ 2 \ln{\dfrac{P_b(\beta \pm \eta \beta \,|\, \nu)}{P_b(\beta, \nu)}} \right]^{-1/2}, \nonumber\\
    \Delta \gamma &=& \eta \gamma \left[ 2 \ln{\dfrac{P_b(\gamma \pm \eta \gamma \,|\, \nu)}{P_b(\gamma, \nu)}} \right]^{-1/2}.
\end{eqnarray}

Here, $\eta=1\%$ sets the scale over which the variation of the penalty function is probed. The error bars shown in Figs.~\ref{fig:phase_diagram_R1}, \ref{fig:exponent_variation}, and \ref{fig:phase_diagram_R2}  are obtained using this method. Of the two estimates in eq. \ref{eq:Error_estimate}, we conservatively report the higher error.

\section{Exponent variation for $R=1$ and phase diagram of $R=2$}
\label{app:Phase_diagram_R2}

The variation of the exponents $\nu$, $\beta/\nu$, and $\gamma/\nu$ as a function of density for the nearest-neighbour case ($R=1$) is shown in Fig.~\ref{fig:exponent_variation}. Due to the singular behavior near $\rho=1$, data points in the range $\rho \in (0.81, 1.19)$ are excluded from the analysis. The averages and standard deviations computed over the remaining densities yield

\[
\begin{aligned}
\nu &= 1.95 \pm 0.18, \\
\beta/\nu &= 0.16 \pm 0.02,\\
\gamma/\nu &= 0.66 \pm 0.03.
\end{aligned}
\]

\begin{figure}
    \includegraphics{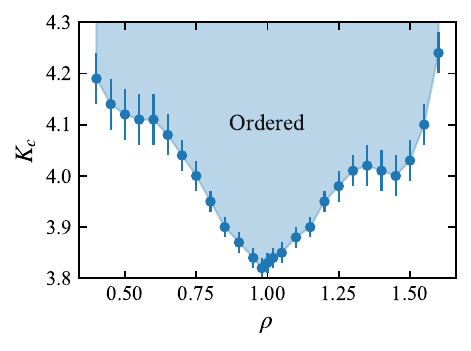}
    \caption{Phase diagram $K_c(\rho)$ for next nearest-neighbour interactions (R = 2)}
    \label{fig:phase_diagram_R2}
\end{figure}

The phase diagram $K_c(\rho)$ exhibits a minimum at $\rho=1$ for all interaction ranges $R > 1$, while this feature is absent for $R=1$. As an illustration, the phase diagram for $R=2$ is shown in Fig.~\ref{fig:phase_diagram_R2}.

\section{Update dependence of the dynamics}
\label{app:parallel_update}

As noted in the main text, the phase behavior depends sensitively on the update scheme. In particular, motility manifests itself differently under asynchronous and synchronous dynamics.
In the asynchronous (random sequential) update, lattice sites are updated one at a time in a random order. This may allow spins to move multiple times within a single time step, leading to enhanced configurational freedom and the possibility of generating defects throughout the bulk.
In contrast, under synchronous (parallel) update, all allowed moves are evaluated and executed simultaneously, such that each spin can move at most once per time step. Starting from a random configuration at $\rho=1$, the system rapidly evolves into configurations with $P=1$, corresponding to locally ordered domains. In this regime, the dynamics becomes effectively confined to domain walls separating regions of opposite orientation.
This can be understood as follows. Each time step consists of a jump followed by a flip. At a domain wall, a spin can hop into a configuration where the flip probability $P$ becomes unity (for example, $P=1$ for $R=1$). A subsequent flip then restores the original configuration at the beginning of the time step, resulting in no net change. As a consequence, bulk dynamics is suppressed, and only spins near domain walls contribute to the evolution.
The remaining dynamics is therefore governed by the stochastic motion of domain walls, which perform an effective random walk. Over long times, this leads the system to one of the absorbing, fully ordered states. In this sense, synchronous dynamics has fewer effective degrees of freedom than asynchronous dynamics and behaves similarly to the non-motile limit.
This argument applies specifically to $\rho=1$. At other densities, the presence of vacancies allows spins to bypass this constraint, preventing complete trapping in absorbing states and enabling sustained fluctuations, analogous to the non-motile case where ordered domains coexist with fluctuating boundaries.

\end{document}